\documentclass[twocolumn,pre]{revtex4}

\usepackage{graphicx}
\usepackage{dcolumn}
\usepackage{bm}

\usepackage{graphics}
\begin{document}

\title{Variable range hopping in thin film with large dielectric constant}

\author{B. I. Shklovskii}
\affiliation{Theoretical Physics Institute, University of
Minnesota, Minneapolis, Minnesota 55455}

\date{\today}

\begin{abstract}
In a film with large dielectric constant $\kappa$ the electric
field of an electron spreads inside the film before exiting the
film at large distances of order of $\kappa d$ ($d$ is the film
width). This leads to the logarithmic Coulomb repulsion between
electrons and modifies the shape of the Coulomb gap in the density
of localized states in a doped film. As a result the variable
range hopping conductivity in such a film has a peculiar
temperature dependence, where the domain of the $\ln \sigma (T)
\propto (T_0/T)^{p}$ dependence, with the index $p \simeq 0.7$, is
sandwiched between the two domains with $p = 1/2$.

\end{abstract}

\maketitle

Variable range hopping (VRH) is the generic mechanism of the low
temperature transport in systems with localized electron states.
When electrons repel each other via the Coulomb potential energy
\begin{equation}
V(r)= \frac{e^{2}}{\kappa r}, \label{V}
\end{equation}
where $\kappa$ is the dielectric constant of the solid the density
of localized states $g(E)$ has the soft Coulomb gap. As a result
the VRH conductivity $\sigma$ obeys the Efros-Shklovskii (ES)
law~\cite{ES,SE84}
\begin{equation}
\sigma(T) = \sigma_0
\exp\left[-\left(\frac{T_{0}}{T}\right)^{1/2}\right] \label{ES}
\end{equation}
both in two-dimensional (2D) and three-dimensional (3D) cases.
Here the characteristic temperature
\begin{equation}
T_{0} = \frac{Ce^2}{\kappa a}, \label{TES}
\end{equation}
(we use the energy units for the temperature $T$), $a$ is the
localization length of electrons and $C$ is the numerical
coefficient~\cite{SE84}.

This paper deals with the situation when the Coulomb interaction
between electrons has a more complicated form. We consider a thin
film with the thickness $d \gg a$ and the large dielectric
constant $\kappa \gg 1$. The film is surrounded by the media with
much smaller dielectric constant, for example just by the air with
$\kappa_{ext} = 1$.

The energy of the Coulomb repulsion of two electrons in the film
was calculated exactly in Ref.~\cite{Keldysh}. Here we present
only asymptotic results with their physics interpretation. Let us
assume that the film is defined by the surfaces $z=\pm d/2$ and
one electron is at $z = x = y = 0$. Then at distances $r \ll d$
its electric field (induction) spreads isotropically. At larger
distances the electric field lines are forced by the large
$\kappa$ to stay inside the film, so that the field spreads along
the radius $\rho = \sqrt{x^2 + y^2}$ of the cylindrical coordinate
system with the same $z$-axis. At $\rho \sim \kappa d$ electric
field lines exit from the film and eventually spread uniformly
over the whole $4\pi$ body angle again.

\begin{figure}[ht]
\begin{center}
\includegraphics[width=0.45\textwidth]{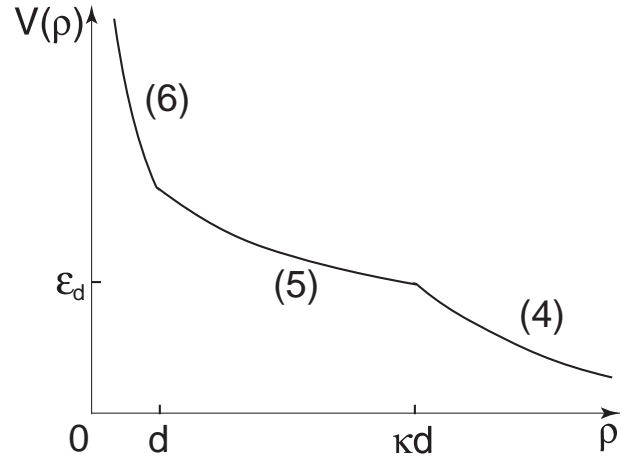}
\end{center}
\caption{A schematic plot of the potential energy of repulsion of
two charges located inside the film with a large dielectric
constant as a function of the distance $\rho$ between electrons.
Equation numbers describing different segments of this plot are
shown next to each segment.} \label{Fig:potential}
\end{figure}

Let us discuss the potential energy of repulsion of two electrons
in the film. At the distance $\rho \gg \kappa d$ two electrons
interact via the "external" Coulomb interaction
\begin{equation}
V(\rho)= \frac{e^{2}}{\rho}. \label{Coulvac}
\end{equation}
On the other hand, in intermediate range of distances $d \ll \rho
\ll \kappa d$
\begin{equation}
V(\rho)= 2\epsilon_d \ln\left(\frac{\kappa d}{\rho}\right),
\label{Coulinter}
\end{equation}
where $\epsilon_d = e^{2}/\kappa d$. Finally at even smaller
distances $\rho \ll d$ we arrive at the the "internal" Coulomb
interaction
\begin{equation}
V(r)= \frac{e^{2}}{\kappa r} + 2\epsilon_d \ln \kappa,
\label{Coulself}
\end{equation}
which differs from Eq.~(\ref{V}) by the logarithmically large
energy accumulated when the second electron is moved from infinity
to $\rho = d$. The resulting potential energy $V(\rho)$ is plotted
as a function of $\rho$ in Fig.~\ref{Fig:potential} in all three
ranges.

Let us assume that the film is uniformly doped by a concentration
of donors $N_D$ and compensated by smaller concentration of
acceptors so that low temperature transport is due to VRH on
donors. How does the two-dimensional electrostatics affect ES law?

This question was first addressed in Ref.~\cite{Larkin} for
another object, a weakly disordered two-dimensional gas, for
example, in a silicon inversion layer. In this case, according to
Ref.~\cite{Abrahams} localized states have an exponentially large
localization length. Large localized states play the dual role in
this theory. First, VRH conductivity is related to long distance
hopping between those of them which are close to the Fermi level.
Second, all other large localized states according to
Ref.~\cite{Larkin} contribute to the large effective dielectric
constant which keeps the electric lines of a charge in the plane
of the two-dimensional gas. In order to calculate VRH conductivity
the authors of Ref.~\cite{Larkin} used an intuitive shortcut
avoiding discussion of the Coulomb gap. The authors found two
ranges of the temperature dependence of the VRH conductivity
$\sigma(T)$. At very low temperatures where the characteristic
length of the hop $r_h$ is much larger than $\kappa d$ they
arrived at ES law with $\kappa = 1$ and and at higher temperature
range where $d \ll r_h \ll \kappa d$ they obtained activated
conductivity.

In this paper, we calculate the VRH conductivity in the framework
of our simpler model, where large $\kappa$ is of the lattice
origin or is a result of close three-dimensional metal-insulator
transition. In this way we avoid the controversial subject of the
2D metal-insulator transition~\cite{Kravchenko}. We follow the
"orthodox" ES logic~\cite{ES,SE84} deriving first the Coulomb gap
and then the conductivity. Let us first formulate our results
moving from high to lower temperatures.

At high temperatures $T \gg \epsilon_{d}a/d$ the VRH hopping
conductivity is determined by the "internal" ES law
\begin{equation}
\sigma(T) = \sigma_0 \exp\left[-\left(\frac{T_{0}}{T}\right)^{1/2}
- \frac{\epsilon_{d}\ln\kappa}{T} \right], \label{internal}
\end{equation}
which at high enough temperatures coincides with Eq.~(\ref{ES}).

It is followed by the range of intermediate temperatures
\begin{equation}
\frac{\epsilon_{d}a}{\kappa d} \ll T \ll \frac{\epsilon_{d}a}{d}
\label{range}
\end{equation}
with "activated" VRH conductivity
\begin{equation}
\sigma(T) = \sigma_0 \exp\left[-\frac{T_{1}(T)}{T}\right],
\label{A}
\end{equation}
where the "activation energy"
\begin{equation}
T_{1}(T) = \epsilon_d\ln\left(\frac{T}{\epsilon_d}\frac{\kappa
d}{a} \right). \label{T1}
\end{equation}
of the intermediate regime is weakly temperature dependent. If one
approximates Eq.~(\ref{A}) as
\begin{equation}
\sigma(T) = \sigma_0
\exp\left[-\left(\frac{T_{p}}{T}\right)^{p}\right], \label{ALPHA}
\end{equation}
the power
\begin{equation}
p = 1 - \frac{1}{\ln(T\kappa d/\epsilon_{d}a)}. \label{p}
\end{equation}
Close to $T= \epsilon_{d}a/d$ we get $p = 1 - 1/\ln \kappa$. For
example, at $\kappa = 40$ one gets $p = 0.7$.

At even smaller temperatures  $T \ll \epsilon_{d}a/\kappa d$ we
arrive at the "external" ES law:
\begin{equation}
\sigma(T) =
\sigma_0\exp\left[-\left(\frac{T_{0}^{ext}}{T}\right)^{1/2}\right],
\label{extES}
\end{equation}
where $T_{0}^{ext} = C e^{2}/a$. Thus, the "activated" VRH
conductivity Eq.~(\ref{A}) is sandwiched between the two different
ES regimes of $\sigma(T)$, the "internal" one,
Eq.~(\ref{internal}), at the high temperature side, and the
"external" one, Eq.~(\ref{extES}), on the low temperature side.
The two low temperature regimes Eq.~(\ref{A}) and
Eq.~(\ref{extES}) are in agreement with Ref.~\cite{Larkin}. The
new high temperature "internal" ES regime exists only if $d \gg
a$.

The experimental literature on the VRH conductivity in thin films
is controversial (see Refs.~\cite{Goldman,Baturina} and references
therein.) How large is the film dielectric constant and how
important is contribution of large localized states~\cite{Larkin}
in most of cases is not clear. On the insulating side of the
superconductor-insulator transition in ultrathin quench-condensed
Ag, Bi, Pb and Pd films~\cite{Goldman} VRH data agree with
Eq.~(\ref{ALPHA}) with $p = 2/3$. On the other hand, for
relatively thick TiN films~\cite{Baturina} the crossover from $p
\simeq 1/2$ to $p \simeq 1$ is observed with the decreasing
temperature. Finding an explanation for this crossover is
challenging~\cite{Baturina} because one would normally expect that
the ES law emerges at the low temperature limit. Our theory shows
that in relatively thick film with the decreasing temperature one
can see the crossover from the "internal" ES law to the activated
transport. This may explain results of Ref.~\cite{Baturina}.

In order to simulate a film with a large dielectric constant one
can make a two-dimensional array of isolated metallic islands
overhanging each other~\cite{Delsing,Fistul}. Although these
arrays were originally designed as arrays of Josephson junctions
they perfectly simulate a large dielectric constant in the normal
state. Indeed, such array keeps electric lines in its plane if the
capacitance between two islands is larger than the capacitance of
each island to the ground. As a result such a normal array in
presence of some disorder should show activated VRH discussed
above.

Before switching to the derivation of our results we would like to
dwell on the related theoretical paper~\cite{Fisher} which deals
with the VRH transport of point like vortexes responsible for the
low temperature resistance of a superconductor film in the
external magnetic film. Two vortexes interact via the logarithmic
potential at small distances, while again at large distances their
interaction follows the "external" Coulomb potential~\cite{Pearl},
so that one could expect to see the two low temperature ranges
discussed above, the "activated" regime and the "external" ES law.
However, Ref.~\cite{Fisher} argues that ES approach is not valid
in this case because "logarithmic interaction grows without bound
with particle separation" and, therefore, "single-particle
energies can not be defined" so that "multi-vortex hopping
dominates the above single-particle effects". The multi-particle
estimate~\cite{Fisher} leads to Eq.~(\ref{ALPHA}) with $2/3 < p <
4/5$. This is close to what we got for $p$ at $\kappa = 40$.

In contrary to the above statements of Ref.~\cite{Fisher}
Fig.~\ref{Fig:potential} clearly shows that the repulsion energy
vanishes at infinity similarly to the standard Coulomb potential
Eq.~(\ref{Coulvac}). Thus, there is no problem to introduce a
single particle energy for a system of localized electrons
interacting with the pairwise potential $V(\rho)$. We can proceed
with the ES argument and study the new shape of the Coulomb gap in
the density of states (DOS) of single particle excitations and
eventually the VRH conductivity. We return to the discussion of
the role of multi-particle processes in the end of this paper.

Below we calculate the zero temperature DOS $g(\epsilon)$
following Refs.~\cite{ES,SE84,ES85}. In the ground state of the
system we define the single electron energy $\epsilon_i$ of an
occupied donor $i$ as the energy necessary to extract an electron
from this state to the state with the energy right at the Fermi
level at infinity. The single electron energy $\epsilon_j$ of an
empty in the ground state donor $j$ is defined as the energy
necessary to bring to it an electron from a state at infinity with
the energy right at the Fermi level $\epsilon_F$. By the
definition of the ground state $\epsilon_j
> \epsilon_F > \epsilon_i$. Another stronger stability condition can be
formulated for each pair of occupied and empty states as follows
\begin{equation}
\epsilon_j - \epsilon_i - V(r_{ij}) > 0. \label{Pair}
\end{equation}
Here $V(r_{ij})$ is the repulsion energy of two electrons on sites
$i$ and $j$ and $-V(r_{ij})$ is the Coulomb energy of attraction
of the electron which moved to the site $j$ by the hole it has
left at the site $i$. In other words, this term describes the
exciton effect. Eq.~(\ref{Pair}) requires that any two states
close in energy to $\epsilon_F$ should be far enough in the space.
This limits the density of states (DOS) close to the Fermi level.
For the Coulomb potential the result for the DOS is known and we
do not repeat derivations from ~\cite{ES,SE84,ES85}, but list the
results.

\begin{figure}[ht]
\begin{center}
\includegraphics[width=0.45\textwidth]{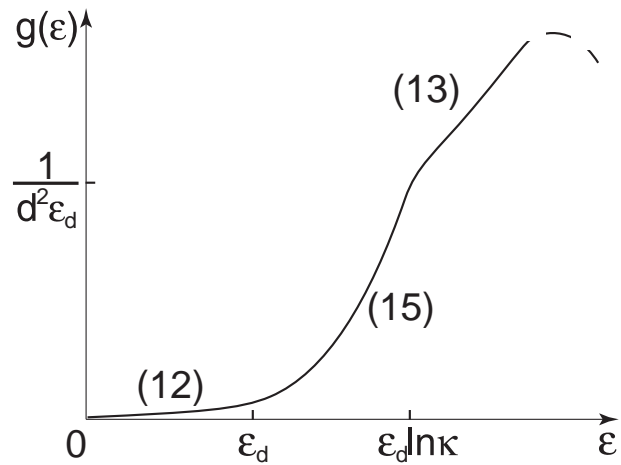}
\end{center}
\caption{Schematic plot of the DOS of localized electrons limited
by Coulomb interaction of electrons as a function of energy.
Equation numbers describing different segments of this plot are
shown next to each segment. Dashed line shows beginning of the DOS
decline at large energies} \label{Fig:CG}
\end{figure}

The "external" Coulomb potential Eq.~(\ref{Coulvac}) limits
two-dimensional density of states at the level
\begin{equation}
g(\epsilon) = \frac{2}{\pi}\frac{\epsilon}{e^4}, \label{2DCG}
\end{equation}
For the "internal" Coulomb potential Eq.~(\ref{Coulself}) we get
\begin{equation}
g(\epsilon) =  \frac{3}{\pi}\frac{\kappa^{3}\epsilon^{2}d}
{e^{6}}, \label{3DCG}
\end{equation}
where the factor $d$ converts the three-dimensional DOS to the
two-dimensional one. Apparently the asymptote Eq.~(\ref{2DCG}) is
valid at large distances, i. e. at small energies $\epsilon$ and
the asymptote Eq.~(\ref{2DCG}) describes small lengths or large
energies. Let us now derive the DOS for the intermediate range of
distances and energies. To this end we have to calculate the
number of states in the band of the width $2\epsilon$ around
$\epsilon_F$. Using Eq.~(\ref{Pair}) and Eq.~(\ref{Coulinter}) we
get $\epsilon_{d}\ln(\kappa d/\rho) < \epsilon$. This means that
in this energy band there is no more than one state in the disc
with radius
\begin{equation}
\rho(\epsilon) = \kappa d \exp
\left[-\frac{\epsilon}{\epsilon_{d}}\right]. \label{radius}
\end{equation}
This leads to the following estimate of the DOS for $\epsilon_d
\ll \epsilon \ll \epsilon_{d}\ln\kappa$:
\begin{equation}
g(\epsilon) \sim \frac{1}{\rho(\epsilon)^{2}\epsilon}=
\frac{1}{(d\kappa)^{2}\epsilon}
\exp\left[\frac{2\epsilon}{\epsilon_{d}} \right]. \label{DOSinter}
\end{equation}
DOS Eq.~(\ref{DOSinter}) matches DOS Eq.~(\ref{2DCG}) at $\epsilon
= \epsilon_{d}$ and Eq.~(\ref{3DCG}) at $\epsilon =
\epsilon_{d}\ln\kappa$ (see Fig.~\ref{Fig:CG})). At large energies
the parabolic range of the Coulomb gap, Eq.~(\ref{3DCG}), is
limited by the total width of the impurity band $\epsilon_{max} =
e^2 N_{D}^{1/3}/\kappa + \epsilon_{d}\ln\kappa$. At $\epsilon \gg
\epsilon_{max}$ the DOS decreases (see the dashed line in
Fig.~\ref{Fig:CG}) similarly to the three-dimensional DOS of the
classical impurity band (see Ch.14 of Ref.~\cite{SE84}).
%

Now we can apply the calculated above DOS in order to estimate the
exponential term of the VRH conductivity of the film at different
temperatures. This estimate closely follows the original Mott's
approach~\cite{Mott}. We define an energy band around the Fermi
level and estimate contribution of this band to the VRH
conductivity. Then we can optimize result with respect of the band
width. This calculations are quite straightforward and we will not
go through them here. Results are already formulated in the
beginning of our paper.

Let us return to the multi-electron effects on ES law. For the
standard Coulomb potential Eq.~(\ref{V}) they were studied in
Refs.~\cite{Efros,Baran,SE84,ES85}. It was shown that they may
change only the coefficient $C$ in Eq.~(\ref{ES}). Here we only
briefly remind what was done. In 3D case a small energy
single-electron excitation strongly interacts with dipole moments
of surrounding compact electron-hole excitations forming together
with them a composite charged multi-electron excitation, the
electronic polaron. Polarons being charged particles obey
stability criterion Eq.~(\ref{Pair}), have the Coulomb gap and
lead to ES law. The only difference is that coefficient $C$ can be
somewhat larger because every hop of a single-electron excitation
in 3D is accompanied by the tunnelling depolarization of many
polaron pairs. A simple estimate showed that the total length of
these small hops is of the order of the length of the main long ES
VRH hop. This gives no more than the factor 2 in the expression
for $C$. In 2D number of dipole excitations in the polaron
atmosphere is of order one and polaron effects provide only small
corrections to $C$.

We estimated similar effects in the framework of the complex
potential of this paper (Eqs.~(\ref{Coulvac}), (\ref{Coulinter}),
(\ref{Coulself})). It turns out that in the low temperature range
polaron effects provide only small corrections to $C$ in Eq.
(\ref{extES}). On the other hand, for the two higher temperature
ranges (Eqs.~(\ref{A}) and (\ref{ES}) the situation is similar to
the 3D case studied in Refs.~\cite{Baran,SE84,ES85}. Namely, at
these temperatures multi-electron effects may add a numerical
coefficient in the exponential. For Eq.~(\ref{A}) this means that
$T_1$ may become twice larger. Thus, multi-electron effects do not
change the power $p$ in Eq.~(\ref{ALPHA}). This conclusion is in
disagreement with Ref.~\cite{Fisher} which assumes that even in 2D
the total length of small hops is much larger than the ES hop and,
therefore, overestimates importance of the multi-electron effects.

Above we discussed only the ohmic transport in a weak electric
field. If the electric field is so strong that $eEa/2 \gg T$ one
can replace $T$ by the effective VRH temperature $eEa/2$ in Eqs.
(\ref{ES}), (\ref{A}), (\ref{ALPHA}) and (\ref{extES}) to obtain
non-ohmic current-voltage characteristics~\cite{Shklovskii} (see
also~\cite{Larkin,Marianer}). For the intermediate activated
regime we then arrive at the current-voltage characteristics $J
\propto J_{0}\exp(-E_0/E)^{p}$, where $p$ is close to unity. This
result is in agreement with the earlier theory~\cite{Nelson} of
the VRH transport of pinned vortexes in superconductors under the
influence of a strong current.

Until now we dealt with a film. Now let us briefly discuss the
similar physics in a long cylindrical nano-wire or nano-rod with
radius $d$ made from a semiconductor with a large dielectric
constant $\kappa \gg 1$, for simplicity, in the air environment.
Here again the electric field of an electron at distances $r \ll
d$ spreads isotropically, then stays inside nano-rod for a
distance $x < \xi = d \kappa^{1/2}$ along the cylinder axis, and
then leaks from the cylinder and eventually spreads isotropically
in the air at large enough $r$. Thus, the potential energy of
repulsion of two electrons again changes from the very short range
"internal" Coulomb interaction Eq.~(\ref{Coulvac}) to the very
long range "external" Coulomb interaction Eq.~(\ref{V}) with the
large intermediate range of $x$, where interaction has the
one-dimensional character. The interaction in this range was
studied~\cite{Kamenev,Zhang} for an ion channel in a lipid
membrane, where the cylindrical pore with radius $d$ is filled by
water with $\kappa = 81$ and is surrounded by lipids with $\kappa
= 2$. Translated to our problem the potential energy of two
electrons located at the nano-rod axis at the intermediate
distance $x$ from each other is well approximated by
\begin{equation}
V(x)= eE_0\xi[\exp(-x/\xi)-1], \label{VNW}
\end{equation}
where $E_0 = 2e/\kappa d^{2}$. (see Sec. VIII of
Ref.~\cite{Zhang}). Using this potential for repulsion of two
electrons in the nano-rod and following the Coulomb gap based
derivation similar to one used above for a film (or the shortcut
approach of Ref.~\cite{Larkin}) one can calculate the temperature
dependence of the VRH conductivity of the nano-rod in the
intermediate temperature range. The result is the strict
activation regime with the temperature independent activation
energy $T_a = eE_{0}\xi$. It is, of course, sandwiched between the
two ES laws, the "internal" one on the high temperature side and
the "external" one Eq.~(\ref{extES}) on the low temperature side.

I am grateful to T. Baturina, A. M. Goldman, A. Kamenev, D. E.
Khmelnitskii and A. I. Larkin for useful discussions.


\end{document}